\documentclass[aps,prb,twocolumn,groupedaddress]{revtex4}
\usepackage{times,xspace}
\usepackage{amsbsy,amssymb,amsmath,bm}
\usepackage{graphicx,color,epsfig,rotate}
\newcommand{\be}{\begin{equation}}
\newcommand{\ee}{\end{equation}}
\newcommand{\bea}{\begin{eqnarray}}
\newcommand{\eea}{\end{eqnarray}}
\newcommand{\vp}{\varphi}
\newcommand{\pr}{\prime}

\newcommand{\cn}{{\rm cn}}
\newcommand{\sech}{{\rm sech}}

\begin{document}
%
\title{Exact Elliptic Compactons in Generalized Korteweg-DeVries Equations}

\author{Fred Cooper$^{\ast,\dag,\ddag}$, Avinash Khare$^{\ast\ast}$, 
Avadh Saxena$^\dag$}

\email{fcooper@nsf.gov}

\affiliation{$^\ast$National Science Foundation, Division of Physics, 
Arlington, VA 22230, USA}

\affiliation{$^\dag$Santa Fe Institute, Santa Fe, NM 87501, USA}
\affiliation{$^\ddag$Theoretical Division, Los Alamos National Laboratory,
Los Alamos, NM 87545, USA}

\email{khare@iopb.res.in}

\affiliation{$^{\ast\ast}$Institute of Physics, Bhubaneswar, Orissa 
751005, India} 

\email{avadh@lanl.gov}


\date{\today}

\begin{abstract}

We derive a general theorem relating the energy, momentum and velocity 
of any solitary wave solution of the generalized KdV equation which 
enables us to relate the amplitude, width, and momentum to the velocity of 
these solutions. We obtain the general condition for linear and Lyapunov 
stability.  We then obtain a two parameter family of exact 
solutions to these equations which include  elliptic and hyper-elliptic 
compacton solutions. For this general family we explicitly verify both 
the theorem and the stability criteria. 
\end{abstract}


\maketitle

In a variety of physical contexts one finds nonlinear field equations 
for which a wide class of solitary wave solutions can exist.  However, 
in many cases one is not able to obtain the solution in a closed form 
and thus it is not very easy to study the stability of such solutions. 
In this Letter we derive a general theorem relating the energy, momentum 
and velocity of any solitary wave of a generalized Korteweg-De Vries 
(KdV) equation of Cooper, Shephard and Sodano (CSS) \cite{CSS}. The 
important point is that this particular generalization of the KdV equation 
is derivable from an Action Principle. Using the theorem, we are able to 
relate the amplitude, width and velocity of any of the solitary wave 
solution even if such a solution is not known in a closed form and also 
study its stability.  Secondly, we obtain a two-parameter family 
of solutions to these equations which include elliptic and hyper-elliptic 
compacton solutions. For this general family we explicitly verify the 
theorem as well as the stability criteria.

Compactons were discovered originally in an extension of the KdV equation 
by Rosenau and Hyman in Ref. \onlinecite{RH}.  Compactons are fundamental 
excitations (i.e., solitary waves with a compact support) of KdV-like 
equations that possess a nonlinear dispersion and collide quasi-elastically. 
They play an important role in pattern formation and emergence of 
nonlinear structures in physical systems.  Other physical contexts in 
which compactons are relevant include fluid dynamics, optical waveguides 
and the field of intrinsic localized modes \cite{kivshar}.  Breather 
compactons \cite{schochet,dinda} and compactons in other nonlinear dispersive 
equations \cite{wazwaz} are also known.  In addition, KdV equations with 
higher power of nonlinearity and dispersion may lead to the phenomema of 
blow-up and collapse.  Therefore, the study of compactons and their 
stability in this class of equations is important in its own right. 

Rosenau and Hyman showed that in a particular generalization of the KdV 
equation, defined by parameters $(m,n)$ (with $m,n$ integers), namely
\be\label{1}
K(m,n) :   u_t + (u^m)_x + (u^n)_{xxx} = 0\,,  
\ee
a new form of solitary wave with compact support is a solution of this 
equation. For the case $m=n$ ($m$ integer) these compactons had the 
property that the width was independent of the amplitude.  In Ref. 
\onlinecite{RH} it was stated that $ K(3,2)$ had an elliptic function 
solution. In a later work, Rosenau  \cite{R} obtained elliptic function 
compactons for the case of $K(4,2)$ and $K(5,3)$.  Phase compactons have 
also been investigated \cite{phase}.  Because the equations of Rosenau 
and Hyman were not equivalent to a Hamiltonian dynamical system, CSS 
\cite{CSS} considered instead a related generalization of the  KdV equation
\bea\label{2}
&&K^*(l,p):  u_t+ u_x u^{l-2} + \alpha[2 u_{xxx}u^p + 4p u^{p-1} u_x 
u_{xx} \nonumber \\    
&& +p(p-1) u^{p-2} (u_x)^3 ] =0\,.
\eea
Equation (\ref{2})  has the property that it is derivable from a 
Lagrangian. CSS showed that the two equations have the same {\it class} 
of solitary wave solutions 
when  $l=m+1$ and $p=n-1$.   Because of this connection, the two parameter
family of new solutions that we will find here will also be solutions of the $K(m,n)$ equation
with slightly different coefficients. 
Cooper and Khare \cite{CK} later showed that these 
equations  with $l=p+2$, i.e. $m=n \le{3}$ 
and $m$ continuous, had compacton solutions of the form 
\be\label{2a}
u(x,t)=A[\cos(\beta y)]^{2/p}\,,
~~ y= x-ct\,,  
\ee
and for all these compactons
also the width was independent of the amplitude. 

The CSS equation has three conserved quantities,  
i.e. the Hamiltonian $H$, momentum $P$ and mass $M$ given 
by \be\label{3}
H= \int \left[ \alpha
u^p (u_{x})^{2} - { {u^l} \over {l(l-1)}} \right]\, dx \,,
\ee
\be\label{4}
P = \frac{1}{2} \int  u^{2}(x,t)\,dx; ~~ M= \int  u(x,t)\,dx\,.  
\ee
On the other hand, the Rosenau-Hyman equation has in general
two conserved quantities (except when $m=n$ in which case there are 
four conservation laws \cite{RH,Dey}) given by
\be\label{5}
M= \int  u(x,t)\,dx; ~~   Q =  \int  u^{n+1}(x,t)\,dx\,. 
\ee
The fact that the CSS equations were derivable from 
an Action Principle allowed CSS to consider time-dependent variational 
approximations based on simple time-dependent trial functions $u^v(x,t)$. 
By using post-gaussian trial wave functions  
$ u^v(x,t)= A(t) \exp{ \left[ - \beta(t) |x+q(t)| ^\gamma \right] } $ it 
was shown that these trial wave functions satisfy the relationship
\be\label{6}
  {\dot q}= r  \frac{H}{P}\,,
\ee
with 
\be\label{7}  r= -\frac{p+l+2}{p+6-l}\,. 
\ee  
Here $\dot{q}$ is the velocity of the compacton.  However, it was 
not known at the time if the result was true in general.  In this Letter 
we will show that this relationship is entirely general and does not depend
on any trial wave function approximation. We will show explicity that this 
relationship can be derived for all solitary wave solutions of the CSS 
equations of the form $u(x,t) = A Z[\beta(x+q(t))]$.  
This relationship will also allow us to relate the amplitude, width and 
velocity of the solitary wave. Having 
a Hamiltonian formulation also simplifies the discussion
of stability, and using 
general arguments we shall prove for the CSS equations that 
 the compacton solutions are stable provided  
$2 < l < p+6$. 

{\it Energy-momentum relation theorem:} 
To derive the relationship between the conserved energy and the conserved 
momentum, the starting point is the action
\be\label{8}
\Gamma = \int L dt\,, 
\ee
where $L$ is given by (note $\phi_{x}=u$) 
\be\label{9}
 L(l,p) = \int \left( \frac{1}{2} \vp_{x} \vp_{t} + { {(\vp_{x})^{l}} \over
{l(l-1)}} - \alpha(\vp_{x})^{p} (\vp_{xx})^{2}  \right) dx\,.
\ee
If we assume the exact solitary wave solution is of the generic form 
\be\label{10}
\phi_x = u =  A Z (\beta(x+q(t)))\,,
\ee
then the  value of the  Hamiltonian for the solitary wave solution 
can be shown to be  
\be\label{11}
H= -C_1 (l) \frac{ A^l}{\beta l(l-1)} + \alpha \beta A^{p+2} C_2(p) , 
\ee
where
\be\label{12}
C_1(l)  = \int Z^l(z) dz; ~~~ C_2 (p)= \int [Z^\prime(z)]^2 Z^p(z) dz \,.
\ee
Since $H$ and $P$ are conserved, 
therefore we can rewrite the parameter $A$ in terms of the conserved momentum $P$ and obtain
\be\label{13}
P= \frac{1}{2} \int dx u^2 = \frac{A^2}{2 \beta} C_5\,,~~
C_5= \int dz Z^2(z)\,. 
\ee
Replacing $A$ by $P$, we now have 
\be\label{14}
H =  -C_3(l) P^{l/2} \beta^{(l-2)/2} + C_4(p)  P^{(p+2)/2} \beta^{(p+4)/2} , 
\ee
where
\be\label{15}
C_3(l) = \frac{C_1 (l)} {l(l-1)}\left[ \frac{2}{C_5}\right]^{l/2}  ;    
~~ C_4 = \alpha C_2(p) \left[ \frac{2}{C_5}\right]^{(p+2)/2} .
\ee

The exact solutions have the property that they are the functions of 
$\beta$ that minimize the Hamiltonian with respect to $\beta$.  Explicit 
examples showing this for both the CSS equation as well as a quintic 
generalization of this equation are found in the appendix of Ref. 
\onlinecite{CHK}.  On using ${\partial H}/{\partial \beta}= 0$, we 
obtain 
\be\label{16}
\beta = P^{\frac{p-l+2}{l-p-6}} \left[ \frac{C_4}{C_3} \frac{p+4}{l-2 }\right]^{2/(l-p-6)}.
\ee
This leads to 
\be\label{17}
H= f(l,p) P^{-r}\,,
\ee
where $r$ is given by Eq. (\ref{6}), 
and
\be\label{18} 
f(l,p) =-\left( \frac{p-l+2}{p+4} \right) C_3(l) \left[ \frac{C_4(p)}{C_3(l)} \frac{p+4}{l-2 }\right]^{(l-2)/(l-p-6)}.
\ee
Hamilton's equation, ${\dot q} = \partial H/\partial P$ now yields the 
relationship as given by Eqs. (\ref{6}) and (8).

Using Eqs. (\ref{6}), (\ref{13}) and (\ref{16}) to (\ref{18}) it is easy 
to show that the momentum $P$, amplitude $A$ and the width parameter 
$\beta$ functionally depend on the  velocity $c$ (note $c=-\dot{q}$) by
\be\label{19}
P \propto  c^{\frac{p+6-l}{2(l-2)}}\,,~~A \propto c^{\frac{1}{l-2}}\,,~~
\beta \propto c^{\frac{l-p-2}{2(l-2)}}\,.
\ee

Several comments are in order.
\begin{enumerate}
\item Notice that when $l=p+2$ then  $\beta$ is independent of the
velocity $c$  and momentum $P$ and hence the amplitude $A$ of the 
solitary wave.
\item Note that the $c$ dependence of the 
amplitude $A$ solely depends on the parameter $l$ and is
independent of the parameter $p$. 
\item In the special case when $p=0$ and $l=d+2$, the CSS equation reduces
to $d$-th order KdV equation. In particular, in that case $d=1$ corresponds
to the KdV equation while $d=2$ corresponds to the modified Korteweg-De 
Vries (mKdV) equation. For this case, a well known exact solution is 
$f(y)=A \sech^{2/d}(\beta y)$, where $\beta=(d/2)\sqrt{v}$ and $2A^d= 
(d+1)(d+2)v$ which indeed is consistent with relation (20).  For that case 
we notice from above that for any $d$, the width parameter $\beta$ varies 
as $c^{1/2}$ while the amplitude $A$ varies as $c^{1/d}$.

\end{enumerate}

{\it Stability of Solutions.}
The stability problem at $l=p+2$ was studied in Ref. \onlinecite{Dey}, using
the results 
of Karpman \cite{Karp1,Karp2}.  Their analysis is in fact also valid 
for arbitrary $l,p$. The result of detailed analysis is that
the criterion for linear stability is equivalent to the condition,
\be
\frac{\partial P}{\partial c} > 0.
\ee
Since for all of our solutions $P \propto c^{(p+6-l)/2(l-2)}$,  it
immediately follows that the solutions are stable provided $2 <l < p+6$.
Analysis of Lyapunov stability following Refs. \onlinecite{Dey,Karp1,Karp2} 
also leads to the same restriction on $p$.

{\it Exact solitary wave and compacton solutions:}
If we assume a solution to (\ref{2}) in the form of a travelling
wave:
 \be\label{20}
u(x,t) =f(y) =f(x-ct)\,,
\ee
we then obtain
\be\label{21}
 c f^{\prime} = f^{\prime} f^{l-2} + \alpha\left(2 f^{\prime \pr \pr}
f^p + 4p f^{p-1} f^{\prime} f^{\prime \pr} +p(p-1) f^{p-2} f^{\prime
3}\right). \ee
Integrating twice we obtain: 
\be\label{22}
{{c} \over {2}} f^2 -{{f^l} \over {l(l-1)}} - \alpha f^{\prime 2}f^p=
K_1 f + K_2.
\ee
For compactons, the integration constants, $K_1$ and $K_2$ are zero.
The general theorem derived above is valid in the case that the integration 
constants $K_1,K_2$ are zero. Unless stated otherwise, throughout this 
paper we shall consider the case when $K_1,K_2$ are both zero. On demanding 
that $ f^{\prime \pr } f^p \rightarrow 0,  f^{\prime 2}f^{p-1} \rightarrow 
0$ at edges where $f\rightarrow 0$, while $f^{\prime}$ is finite at edges 
gives us the following bounds on $l$ and $p:l>1,0<p \le 2, p \le l$.

It is worth noting that Eq. (\ref{22}) is very similar to the equation obtained for the Rosenau-Hyman case:
\be\label{23}
(f^\prime)^2 =  \frac{2v}{n(n+1)} f^{3-n} 
- \frac {2}{n(n+m)} f^{m-n+2}\,. 
\ee
Thus, we see that with $l=m+1$ and $p=n-1$ the equations for finding solutions are identical in 
form, with only differing coefficients.  Therefore we expect to find similar solutions to the
two sets of equations.  

Let us now look at the various different compacton solutions to Eq. 
(\ref{22}). For the particular case $l=p+2$  ($m=n$),  Cooper and Khare 
\cite{CK} were able to show that the CSS equation has solutions of the form
(\ref{2a}) and that for all these compacton solutions the width is 
independent of the amplitude. 
We now show that if instead $l=2p+2$ then one gets a one parameter family of
elliptic compacton solutions. In particular, it is easily shown that
\be\label{24}
f=A \cn^{\gamma} (\beta y,k^2=1/2)\,,~~ 
\ee
for  
\be
-K(k^2=1/2) \le \beta y \le  K(k^2=1/2)\,, 
\ee
and zero elsewhere is an exact elliptic compacton solutions to the field 
Eq. (23) provided
\bea\label{25} 
&&\gamma = 2/p\,,~~l=2p+2\,, ~~ 
A^{2p}=c(p+1)(2p+1)\,, \nonumber \\
&&\beta^4=\frac{cp^4}{16\alpha^2 (p+1)(2p+1)}\,.
\eea
The $(m,n)=(3,2)$ and $(m,n)=(5,3)$ are  two special
cases of solution (\ref{24}) with  $\gamma=1,2$.  Here $cn(y,k)$ is a 
Jacobi elliptic function and $K(k)$ denotes the complete elliptic 
integral of the first kind with modulus $k$.  

If instead $l=3p+2$, then we obtain a one parameter family of elliptic 
solutions of the form
\be\label{26}
f=A\left[\frac{1-\cn(2(3)^{1/4}\beta y,k^2)}{(1+\sqrt{3})+(\sqrt{3}-1) 
\cn(2(3)^{1/4} \beta y,k^2)}\right]^{\gamma}\, , 
\ee 
with the modulus  
\be\label{27} 
k^2=\frac{1}{2}-\frac{\sqrt{3}}{4}\, .
\ee    
Here
\bea\label{28}
&&\gamma=1/p\,,~~l=3p+2\,,~~2A^{3p}= c(3p+1)(3p+2)\,, \nonumber \\
&&\beta^6 = \frac{c^2 p^6}{256\alpha^3 (3p+1)(3p+2)}\,.
\eea
This is a compacton solution in the range 
\be\label{29}
0 \leq w=2 (3)^{1/4} \xi \leq 4 K\left(k^2 =\frac{1}{2}-\frac{\sqrt{3}}{4} 
\right).
\ee
and zero elsewhere, where $\xi=\beta y$.
The way this solution is obtained is by starting with the ansatz
$f=AZ^{a}(\xi)$ and demanding that $Z$ satisfies the differential
equation
$({dZ}/{d\xi})^2= 1-Z^6$.
The above integral can be evaluated by converting the differential equation
into the standard differential equation for the Weierstrass elliptic 
function \cite{GR} ${\cal P}(y,k)$. On simplifying, we obtain the explicit 
solution as given by Eq. (28). The $(m,n)=(4,2)$ solution
of Ref. \onlinecite{R} is a special case of our general solution (\ref{26}) with $p=1$.

We now show that all the above solutions are in fact special cases of the 
two parameter ($a,t$) family of solutions obtained by assuming
\be\label{30}
 f= A Z^a (\xi=\beta y)\,,
\ee
and demanding that
\be\label{31}
(Z^\prime)^2 = 1- Z^{2t}\,. 
\ee
We find the conditions:
\bea
&& l=pt+2; ~~ a=2/p\,,~~
2A^{pt} = (pt+2)(pt+1)c\, , \nonumber \\ 
&&\beta^{2t}=\frac{c^{t-1}p^{t}}{\alpha^t 2^{3t-1}(pt+1)(pt+2)}\,.
\eea
It may be noted that here $l,p,t$ are all continuous (i.e. real) parameters. 
The various solutions discussed above correspond to $t=1,2,3$.  In fact 
an explicit compacton solution can also be found in the case $t=3/2$.  
The solution for $t=3/2$ is essentially the same as that for $t=3$, Eq. 
(29), but with Eq. (30) replaced by the complementary modulus $k^2= 
\frac{1}{2}+\frac{\sqrt{3}}{4}$. For $t>3$, the compacton solutions are 
related to hyper-elliptic functions \cite{GR}. 

It is interesting to note that even though none of the hyper-elliptic solutions
can be obtained in an analytic form $(t > 3$), still their momentum and
energy can be obtained analytically.
In particular, on using Eq. (\ref{31}) in the expressions for $H,P$ as given 
by Eqs. (4) and (\ref{4}) it is easily shown that 
\be
H= -\frac{A^2 c}{2 \beta t } \frac{(6+p-l)}{(l+p+2)} {\rm B} 
\left(\frac{p+4}{2pt}, \frac{1}{2}\right), 
\ee
\be
P =   \frac{A^2 }{2 \beta t } {\rm B} \left(\frac{p+4}{2pt},\frac{1}{2}\right), 
\ee
where $B$ denotes the Beta function \cite{GR}.  For all these solutions 
with $p, l$ continuous (i.e. real) variables, the relationship $H/P=c/r$ is 
always satisfied.

For the hyper-elliptic compacton solutions we have that $l$, $p$ and $t$ 
are related by $l=pt+2$ so that the general stability criterion can be 
rewritten as $0<pt<p+4$, or for $t>1$  
\be
0 < p(t-1) < 4. 
\ee
The requirement for non-singular solutions is that $0 < p \leq 2$.  This means
for $t \leq 3$, compactons with arbitrary $p$ in the allowed range are
linearly stable while when $t  > 3$ , the compactons are stable only for
$0 < p < 4/(t-1)$.
Analysis of Lyapunov stability following Refs. \onlinecite{Dey,Karp1,Karp2} 
also leads to the same restrictions on $p$.

From the stability analysis
 we also recover the well known result that the higher order KdV
equations, characterized by $p=0,l=d+2$, have stable soliton solutions 
provided $d<4$ while one has unstable soliton solutions in the case 
\cite{rus,kuz1,kuz2} $d>4$.

{\it General remarks.} 
In conclusion, we have obtained explicit exact compacton solutions in 
terms of Jacobi elliptic functions which exist at particular values 
of the elliptic modulus $k$.  Note that at $k=0$ these solutions do not 
exist.  In addition, we derived a quite general theorem that relates 
the energy momentum and velocity of solitary wave solutions without 
knowing the explicit form of the solution.  We note that the above 
analysis should also hold for the generalized quintic KdV equation. 
We also notice that the radial part of the generalized nonlinear 
Schr\"odinger equation with nonlinear power $\kappa$ 
\be 
i\frac{\partial\psi}{\partial t}+\nabla^2\psi+g|\psi*\psi|^{\kappa} 
=0 
\ee 
obeys the same equation as 
(24) with $p=0$ and $l=2\kappa+2$, so that the stability analysis of 
these two problems are related.  We believe that it should be 
possible to derive a similar theorem in other nonlinear
systems.

Finally, as a byproduct of our results, we are able to obtain analytic 
expressions for $K(k)$ and the complete elliptic integral of the second 
kind, $E(k)$, at $k^2=1/2-\sqrt{3}/4$ (see Eq. 30) or at $k=\sin(\pi/12)$. 
Specifically, 
\be 
K\left(k=\sin\left(\frac{\pi}{12}\right)=\frac{\sqrt{3}-1}{2\sqrt{2}} 
\right)=\frac{3^{1/4}\sqrt{\pi}\Gamma(1/6)} 
{6\Gamma(2/3)}\,,   
\ee 
where $\Gamma(a)$ denotes the Gamma function \cite{GR}.  In addition, 
at $k=\sin(\pi/12)$ using the relations $K'=\sqrt{3}K$ and  
\be 
E=\frac{\pi\sqrt{3}}{12K}+\sqrt{\frac{2}{3}}~k'K , ~~~ 
E'=\frac{\pi\sqrt{3}}{4K'}+\sqrt{\frac{2}{3}}~k'K' , 
\ee  
we also have the explicit analytic expressions for $E$ and the two 
complete elliptic integrals with complementary modulus $k'=\sqrt{1-k^2} 
=\cos(\pi/12)$, namely $K'$ and $E'$. 
 
F.C. would like to thank the Santa Fe Institute and A.K. would like to 
thank the Center for Nonlinear Studies and Theoretical Division at 
LANL for their hospitality during the completion of this work. We would 
also like to thank the U.S. DOE and the NSF for their partial support 
of this work.

\end{document}